\documentclass[aps,prd,twocolumn,showpacs,nofootinbib]{revtex4-1}

\usepackage[latin1]{inputenc}
\usepackage{pict2e}
\usepackage{latexsym}
\usepackage{amsmath,amsfonts}
\usepackage{amsbsy}
\usepackage{mathrsfs}
\usepackage{color}

\usepackage{psfrag}

\usepackage{enumerate}

\usepackage{amsmath,amssymb,calc,amsfonts}
\usepackage{latexsym}

\usepackage{graphicx,calc,epsfig}
\usepackage{hyperref} 
\usepackage{float}

\def\ut#1{\rlap{\lower1ex\hbox{$\sim$}}#1{}}

\newcommand{\R}{\mathbb{R}}
\newcommand{\be}{\nopagebreak[3]\begin{equation}}
\newcommand{\ee}{\end{equation}}
\newcommand{\ba}{\nopagebreak[3]\begin{eqnarray}}
\newcommand{\ea}{\end{eqnarray}}
\DeclareFontFamily{U}{rsfs}{}         
\DeclareFontShape{U}{rsfs}{m}{n}{<5> rsfs5 <6><7> rsfs7          %
  <8><9><10><10.95><12><14.4><17.28><20.74><24.88> rsfs10}{}     %
\DeclareMathAlphabet{\mathfs}{U}{rsfs}{m}{n}                     %
\newcommand{\mfs}[1]{\mathfs {#1}}                               %
\newcommand{\n}{{\nonumber}}

\newcommand{\sM}{{\mfs M}}

\newcommand{\sI}{{\mfs I}}\newcommand{\sO}{{\mfs O}}

\def\i{i}

\def\pb#1{\rlap{\lower1.5ex\hbox{$\longleftarrow$}}{#1}}
\def\dpb#1{\rlap{\lower1.5ex\hbox{$\Longleftarrow$}}{#1}}
\def\spb#1{\rlap{\lower1.5ex\hbox{$\leftarrow$}}{#1}}
\def\sdpb#1{\rlap{\lower1.5ex\hbox{$\Leftarrow$}}{#1}}

\definecolor{blue}{rgb}{0,0,1}
\definecolor{green}{rgb}{0,1,0}
\definecolor{red}{rgb}{1,0,0}
\definecolor{vio}{rgb}{1,0,1}
\definecolor{ama}{rgb}{1,1,0}

\begin{document}

%
%



\title{Smooth null hypersurfaces near the horizon in the presence of tails}

\date{\today}

\author{Carlos Kozameh$^1$, Osvaldo Moreschi$^1$ and Alejandro Perez$^2$}



\affiliation{$^1$ FaMAF, UNC, Instituto de F\'\i{}sica Enrique Gaviola (IFEG), CONICET, \\
Ciudad Universitaria, (5000) C\'ordoba, Argentina. }

\affiliation{$^2$Centre de Physique Th\'eorique\footnote{Unit\'e
Mixte de Recherche (UMR 6207) du CNRS et des Universit\'es
Aix-Marseille I, Aix-Marseille II, et du Sud Toulon-Var; laboratoire
afili\'e \`a la FRUMAM (FR 2291)}, Campus de Luminy, 13288
Marseille, France.}

\begin{abstract}
We show that the power-law decay modes found in linear perturbations of Schwarzschild black holes, 
generally called tails,
do not produce caustics on a naturally defined family of null surfaces 
in the neighbourhood of $i^{+}$ of a black hole horizon.
\end{abstract}


\maketitle
\section{Introduction}

\begin{figure}[h!]
\centering
\includegraphics[clip,width=0.45\textwidth]{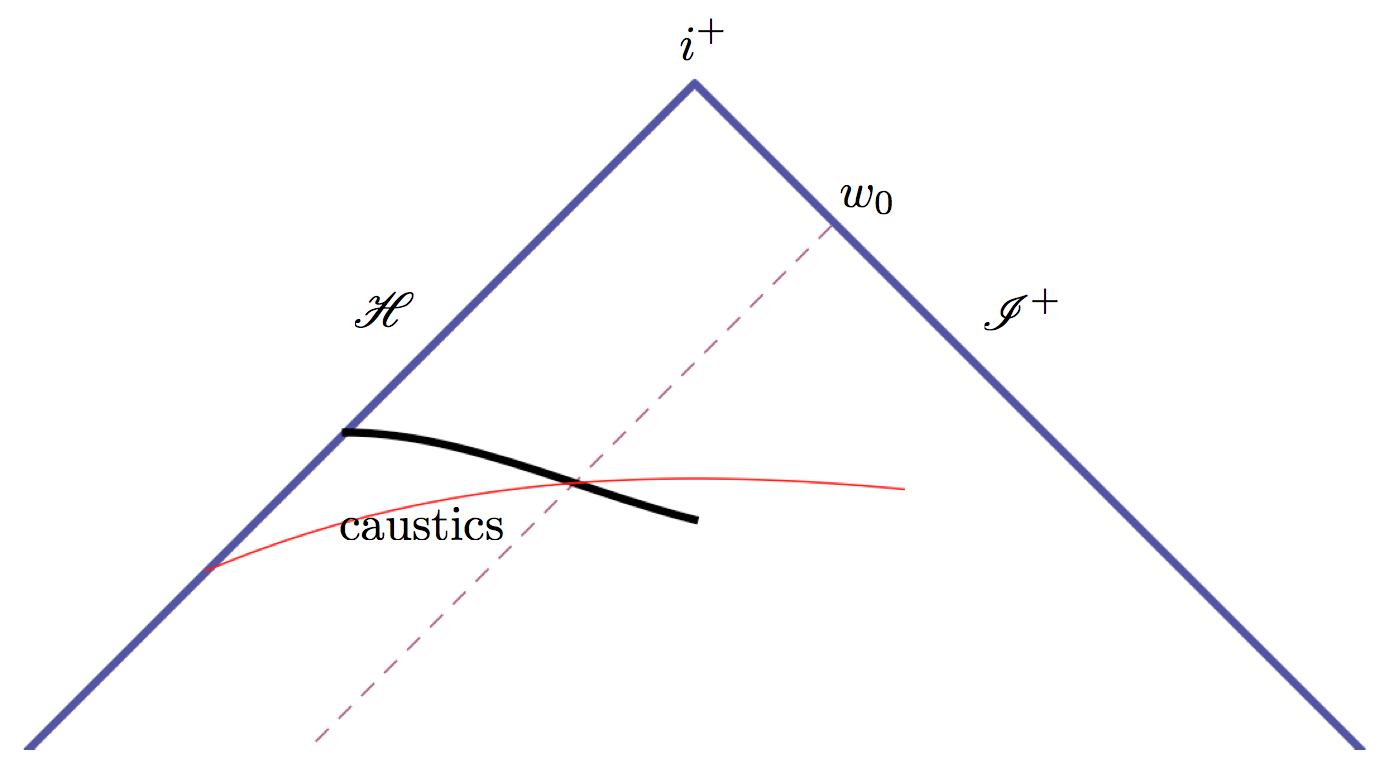}
\caption{Conformal diagram representing the gravitational collapse 
producing a {\em solitary black hole}. There is $w_0<0$ such that for $w_0<w<0$
there is a caustic free neighbourhood around $i^{+}$ containing a portion of the horizon $H$
and $\sI^{+}$, if the spacetime decays 
towards its final stationary state sufficiently rapidly.  
}
\label{SBH}
\end{figure}

A new framework for the dynamical description of the late phase of gravitational collapse
has been recently proposed \cite{Kozameh10,Kozameh11}. In this framework one introduces physical null 
coordinates based on the assumption that a suitable family of null surfaces are caustic free
in a neighbourhood of timelike infinity containing a portion of the black hole horizon $H$ and 
future null infinity $\sI^{+}$.
We consider an asymptotically flat spacetime at future null infinity $(\sM,g_{ab})$ containing a black hole.
Its conformal diagram is depicted in Figure \ref{SBH}. In  the past of an open set of future null infinity 
($\sI^+$)---defined by those points for which
their Bondi%
\footnote{A Bondi retarded time $u$ is such that the sections $u$=constant at $\sI^{+}$ (referred to as {\em Bondi cuts}) have an intrinsic metric given by (minus) the metric of the unit sphere. }
 retarded time $u$ 
 is in the range $u \in (u_0, \infty)$---we require the existence of a regular null function $w$ 
such that: $w=0$ at the horizon $H$,  and $w<0$ in the region of interest.

Choosing a Bondi coordinate $u$ that coincides with 
the center of mass Bondi cuts\cite{Kozameh10,Moreschi04} in the regime $u\to \infty$ limit,
we can uniquely fix the function $w$,
if we assume the topology of the black hole (BH) event horizon 
$H$ is $S^2\times \R$ in that region.
Thus,
there exists a smooth null function $w=w(u)$ (unique up to constant scaling 
 in the region where one neglects $O(w^2)$ effects) such that
$w=0$ at the horizon $H$, $\dot w \equiv \frac{dw}{du}>0$, $w < 0$ for all $u$, and $\lim\limits_{u\rightarrow \infty} w = 0$.

This construction is precisely described in \cite{Kozameh12}, where spacetimes satisfying this assumption are defined 
as {\em solitary black holes} (SBBs).
In a few lines, the null geodesic congruence defined by $\tilde \ell=du$ allows for the introduction of 
an affine parameter  $r$  used as a radial coordinate which is fixed by the requirement that it coincides asymptotically with
the luminosity distance (see equation (\ref{lumy}) below for a precise statement of this condition). 
The surfaces $(r,u)=$constant are spheres which inherit natural spherical coordinates defined in the Bondi 
cuts at $\sI^{+}$ which label null rays of the congruence $\tilde \ell$. All this provides a coordinate system $(u,r,\theta,\phi)$ in the 
exterior of the BH horizon.

However, the above coordinate system is not well behaved near the horizon ($u\to \infty$).
A good coordinate system can be constructed if one follows similar lines as above but describing the null geodesic 
congruence instead in terms of $\ell=dw$. One can introduce an affine parameter $y$ along $\ell$ and fix the ambiguity in such 
choice by requiring that the spheres $(w,y)=$constant coincide with the $(u,r)=$constant in the interior of the spacetime. Thus the angular coordinates 
can be defined exactly in the same way as in the previous paragraph. With this one obtains the following relationship between the affine parameters
$r$ and $y$:
\be\label{erre}
r=\dot w y+r_0(w),
\ee
where $\dot w\equiv (dw/du)$. The coordinate $y$ will be used in what follows.

Under mild regularity conditions SBHs are then shown to  posses a smooth global vector field 
\begin{equation}
\chi \equiv \frac{\partial}{\partial u},
\end{equation}
which is a null geodesic generator at $\sI^{+}$ and  a null geodesic generator of the horizon $H$.
Moreover, at the horizon $H$, $\chi$ satisfies the equation,
\begin{equation}\nonumber
\chi^{a}\nabla_{a}\chi^{b} \equiv \kappa \chi^{b} ;
\end{equation}
where $\kappa$ is a  generalized surface gravity.
Finally, one can show that \be\label{main} 
{
w(u)=-\exp{(-\kappa (u-u_0))}+\sO(\exp{(2au)}) }  ,
\ee
where $\exp(-\kappa u_0)$ is the rescaling freedom associated with the choice of origin for
the Bondi retarded time $u$.  The last equation is a generalization of the  Kruskal coordinate
transformation that appears in Schwarzschild and Kerr geometries.

SBHs have thus remarkable global features that can provide additional structure 
in the study of the late phase of gravitational collapse in terms of the full non-linear 
regime of Einstein's equations. The key question is whether the assumption of the existence of the physical null 
function $w(u)$ is too restrictive admitting only situations of little physical interest. 
The whole formalism rests on the assumption that there are no caustics,
in a small enough neighbourhood of $i^{+}$, in the congruence of generators of the
null surfaces $u=constant$ as one goes from $\sI^{+}$ towards the past,
containing a final portion of $H$ and $\sI^{+}$.
We will see that this problem does not appear in the final phase collapse provided by 
the scenario developed in the framework of linear perturbations of stationary BH spacetimes.
This provides a strong indication that our assumptions  are mild enough to admit physically interesting
situations.

As we have seen, there are two coordinates and null tetrad system that one can use near the black hole;
the tilde system that comes from the asymptotic description of the black hole, and the un-tilde
system that it is regular at the horizon. In what follows we work in the tilde system, in order
to make contact with calculations of other authors.

We will study in detail the behavior of the optical scalars $(\tilde\rho, \tilde \sigma)$
which depend explicitly on the incoming gravitational radiation $\tilde\Psi_0$, the
in-falling of matter $\tilde{\Phi}_{00}$, and implicitly in the outgoing gravitational
radiation field $\tilde\Psi_4^0$. Since we center the discussion in the behaviour of the
optical scalars in a neighborhood of the horizon, we will concentrate on the dependence on the fields
$\tilde\Psi_0$ and $\tilde{\Phi}_{00}$ directly. 
In this work we will consider  whether fields with typical tail 
behaviour\cite{Gundlach94,Dafermos:2005yw} are admitted in our setting.

In figure \ref{SBH} it is shown the horizon $H$, future null infinity $\sI^{+}$, timelike infinity $i^{+}$ and
the region of interest that is for $w > w_0$ and $y>y_0$; where the hypersurface $w_0$ is denoted by a dash line
and $y_0$ by a thick black line. It is important for the study to understand the behaviour of the fields
in a neighbourhood of the horizon but for finite values of $y$.

In \cite{Kozameh10} we point out that $\Psi_0=\dot w^2 \tilde \Psi_0$ and $\Phi_{00}=\dot w^2 \tilde \Phi_{00}$ 
must go as $y^{-3}$ on the horizon in order for the area of the horizon 
to have an asymptotic finite value, in the limit $y\to \infty$.


Since we have not found in the literature a general discussion regarding the behaviour of $\tilde \Psi_0$ 
in the same asymptotic region near the horizon; 
from our knowledge on the behaviour of $\Psi_0$ at the horizon and the behaviour of 
 $\tilde \Psi_0$ in the asymptotic region, we will assume the worst possible scenario.
At the horizon we know that $\Psi_0$ can behave as $y^{-3}$, 
and for $w \neq 0$ this 
means\footnote{Note that at the horizon, i.e. when $w=0$, the relation between $v$ and $y$ is 
logarithmic. However, our study only concerns the region $w \neq 0$.}
that $y^{-3}\sim \dot w^3 v^{-3}$.
In the asymptotic region, for $r\to\infty$ one knows that  $\tilde \Psi_0$ behaves as $r^{-5}$;
which means $v^{-5}$. So we will assume the worst admissible behaviour in the region of interest;
which is to take $\tilde \Psi_0 \sim v^{-3}$.

We will show in Section \ref{tails} that the late time behaviour predicted by the study of matter fields on the 
Schwarzschild background imply
that $\tilde \Phi_{00}$ going as $v^{-4}$, i.e.; even faster than required by the above general argument. 
Thus, in what follows we assume
\be\label{pipi}
\tilde \Psi_0\sim v^{-3}\quad \text{and} \quad  \tilde\Phi_{00}\sim v^{-4}.
\ee  

The article is organized as follows. In the following section we analyze the conditions for caustic formation.
In order to illustrate a way in which we could easily violate our assumptions---and in order to provide a 
clear-cut intuition---we will provide 
what is probably the simplest manner in which one can introduce caustics that invalidate our construction 
in Section \ref{dust}.
We also argue in that section why such possibility is not of interest in the study of the final phase of 
gravitational collapse.
In Section \ref{tails} we briefly review the results of \cite{Gundlach94}.
In Section  \ref{tailseint} we show that the late time behaviour of gravitational collapse expected from the 
linear perturbation technology is admited by our assumptions.

\section{The caustic freeness conditions}

The optical scalars equations can be expressed as
\begin{equation}\label{eq:thornrho-l}
\frac{\partial {\tilde \rho}}{\partial r}
=
{\tilde \rho} ^{2}
+{\tilde \sigma} \, \bar{\tilde \sigma}
+\tilde{\Phi}_{00} ,
\end{equation}
\begin{equation}\label{eq:thornsigma-l}
\frac{\partial {\tilde \sigma}}{\partial r}
=
2 {\tilde \rho} \, {\tilde \sigma}
+\tilde\Psi_0
,
\end{equation}
where $r$ is an affine parameter along the null geodesics $\tilde\ell=\partial_r$ which we will take to coincide 
with the luminocity distance as one approaches future null infinity along the geodesics.

Let us concentrate in the behavior of ${\tilde \rho}$ and study the points in which it has a
divergent behavior: {\em caustics}.
Then one can write  (\ref{eq:thornrho-l})  as
\begin{equation}\label{eq:thornrho-l2}
-\frac{\partial }{\partial r}\left( \frac{1}{{\tilde \rho}}\right) = 
\frac{1}{{\tilde \rho} ^{2}}\frac{\partial {\tilde \rho}}{\partial r}
=
1
+
\frac{{\tilde \sigma} \, \bar{\tilde \sigma} +\tilde{\Phi}_{00}}{{\tilde \rho} ^{2}}
 .
\end{equation}
The previous equation is equivalent to the following integral equation
\begin{equation}\label{inte}
- \frac{1}{{\tilde \rho}(r_\infty)} +\frac{1}{{\tilde \rho}(r)}
=
r_\infty - r
+
\int_r^{r_\infty}
\frac{{\tilde \sigma} \, \bar{\tilde \sigma} +\tilde{\Phi}_{00}}{{\tilde \rho} ^{2}}
dr'
 .
\end{equation}
We would like to study this equation in the limit $r_\infty\to\infty$.
Now, because we have chosen $r$ to agree with the notion of luminocity distance in the large $r$ limit (which is possible if the spacetime is asymptotically flat at future null infinity),
one has that 
\be
\tilde \rho=-\frac{1}{r}(1+\frac{\tilde \rho_1}{r^2}+ O(r^{-3})) 
\ee
this implies that
\be
\frac{1}{\tilde \rho}=-\frac{r}{(1+\frac{\tilde \rho_1}{r^2}+O(r^{-3}))}=-r + O(r^{-1})
\ee
The previous equation implies that 
\be\label{lumy}
\lim_{r_\infty \to \infty} \; \left(\frac{1}{\tilde \rho(r_\infty)} +  r_\infty \right) = 0 .
\ee
In fact the previous condition is the precise definition of $r$ being asymptotically 
the luminocity distance.
Therefore, equation (\ref{inte}) implies
\begin{equation}\label{roro}
{\tilde \rho}(r)
=-\frac{1}{r 
-
\int_r^{\infty}
\frac{{\tilde \sigma} \, \bar{\tilde \sigma} +\tilde{\Phi}_{00}}{{\tilde \rho} ^{2}}
dr'}.
\end{equation}
Thus the condition that caustics appear at $r=r_c$ becomes simply
\be\label{caustics}
\int_{r_c}^{\infty}
\frac{{\tilde \sigma} \, \bar{\tilde \sigma} +\tilde{\Phi}_{00}}{{\tilde \rho} ^{2}}
dr=r_c
\ee
From the previous equation and from the positivity of the integrand involved one can conclude that the condition  
\be\label{caustics2}
\int_{r_1}^{\infty}
\frac{{\tilde \sigma} \, \bar{\tilde \sigma} +\tilde{\Phi}_{00}}{{\tilde \rho} ^{2}}
dr \le r_1
\ee
guaranties the absence of caustics in the interval $r\in(r_c<r_1,\infty)$.
However, the presence of the expansion itself in the previous equation makes this condition a bit cumbersome.
We can turn the previous criterion for the absence of caustics into a
sufficient condition of a simpler and more useful form thanks to the validity 
of the following statement.
\vskip.2cm
\noindent {\bf Lemma:} In the caustic free region $r\in (r_c,\infty)$ the following inequality holds
\be
|\tilde \rho|\ge \frac{1}{r} . 
\ee
The proof follows directly from equation (\ref{roro}), the fact that $0\le {\tilde \sigma} \, \bar{\tilde \sigma} +\tilde{\Phi}_{00}$, and the fact that $r\in  (r_c,\infty)$.
More explicitly, 
\ba
|\tilde \rho|&\ge& \frac{1}{r}   \Longleftrightarrow -\tilde \rho\ge \frac{1}{r}\n \\
&\Longleftrightarrow& r\ge r-\int_{r}^{\infty}
\frac{{\tilde \sigma} \, \bar{\tilde \sigma} +\tilde{\Phi}_{00}}{{\tilde \rho} ^{2}}
dr' \ge 0,
\ea
where we have used the positivity stated in the last inequality which follows from the
condition that $r\in (r_c,\infty)$. The condition that one is in the caustic free region is essential $\square$.
\vskip.2cm

Using the previous result we can write a sufficient condition for the 
non existence of caustics in the interval $r\in (r_1,\infty)$ as follows
\be\label{caustics1}
\int_{r_1}^{\infty}
({{\tilde \sigma} \, \bar{\tilde \sigma} +\tilde{\Phi}_{00}} ) r^2
dr \le r_1.
\ee
The previous condition on the strength of  ${{\tilde \sigma} \, \bar{\tilde \sigma} +\tilde{\Phi}_{00}} $ is clearly
stronger than (\ref{caustics2}). This is why in contrast to the latter this is a sufficient condition (its violation may not imply that there are caustics in $(r_c<r_1,\infty)$).
However, if (\ref{caustics1}) is satisfied then we can assure that there are no caustics in the region of interest. This last condition will be central in the 
proof of our main result in the following section.


\section{Dust}\label{dust}

In this section we show that a grain of sand can destroy our construction.
This simple example will provide intuition on what the nature of our problem is.
At the same time we shall see by the end of this section that this example is physically irrelevant for the
physical situation that one would like to describe in our framework.
 
We can model  a grain of sand (or a planet) at some coordinate $r_d(w)>r_H$ outside de BH horizon
by a Ricci spinor component
\be
\Phi_{00}=\frac{\epsilon}{r_H}\delta(r-r_d(w)),
\ee 
where $\epsilon$ is a dimensionless parameter measuring the strength of the dust particle.
For the next discussion it is enough to use the fact that $\tilde\sigma$ is bounded by $\frac{\alpha}{r^2}$,
in the asymptotic region,
for an appropriate $\alpha$; however for simplicity we will assume next that $\tilde\sigma=0$.
This will not change the qualitative aspects of the discussion.
Then, condition (\ref{caustics}) becomes
\be\label{prima}
 \epsilon \frac{r_d^2}{r_H}
\le r_c.
\ee
In order to define the region where we will proof that there are no caustics we need to recall that
\ba
&& r=\dot w y+r_{H}\n \\
&&=-\frac{w}{2 r_H} y+r_H.
\ea
In order to show that there is a caustic free region around $i^{+}$, containing both a portion $\sI^{+}$ and the
black hole horizon, it is sufficient to show that for a given $y_1$ there exist an $w_0\le 0$ such that for all $w>w_0$ there are no caustics in the region
\be r\in (-\frac{w}{2 r_H} y_1+r_H, \infty).\ee
Without loss of generality, and in order to simplify some expressions, we take $y_1=2 r^2_H$ from now on. 
The region of interest now becomes $r\in (r_H (1-w), \infty)$. 
Thus, from (\ref{prima}), the caustic free condition becomes
\be
\epsilon r_d 
\le r_H(1-w).
\ee
Conversely, the previous equation tells us that it is very easy to introduce caustics 
that would completely invalidate the construction; it suffices to take a dust particle that is sufficiently far away and sufficiently strong.
In particular if we take $\label{sand} \epsilon r_d >r_H(1-w)$ then there will be a caustic line that goes all the way up to $i^{+}$.

Therefore, we have shown that our construction breaks down if a suitable grain of dust is brought in.
Is this a serious problem? We now argue that it is not; as the above situation
bears not interest for the study of the physics of gravitational collapse we plan  to study.
The reason is that the problematic grain of sand (which could also model a planet or a star)
must stay outside the black hole $r_d>r_H$ for all $w$; hence, it is a compact object that is never absorbed
by the BH and follows a timelike trajectory all the way up to $i^{+}$. The only physically acceptable
possibility is then that the object is not gravitationally bound to the BH. Such possibility is of course 
physically viable but it introduces an irrelevant complication to the problem of studying the final stage 
of gravitational collapse. Therefore, it is advisable  that our definition of SBH rules out such
situation by assumption.

\section{Tails}\label{tails}

Gundlach, Price and Pullin \cite{Gundlach94}  have shown that the spherical harmonic $\ell$ mode of 
a scalar field $\phi_0^{\ell}$ satisfying the wave equation on a Schwarzschild
background in the late time behaviour for $u\to \infty$ is
\be
\phi_0^{\ell}=\frac{\Upsilon_0}{v^{P+2\ell+1}},
\ee
where $\Upsilon_0$ is a constant, where $P=1,2$. 
If such scalar field is used as matter source for Einsteins equation then it produces a Ricci
scalar $\tilde\Phi_{00}$ whose late time behaviour is
\be
\tilde \Phi_{00} \approx \frac{1}{v^4}+O(v^{-5}) .
\ee
As explained in expression (\ref{pipi}),
 $\tilde \Psi_0$ goes like $1/v^3$. From the optical equations it follows that $\tilde\sigma$ goes like $1/v^2$.
This means that the late time behaviour of the integrand in (\ref{caustics})
can be expressed as:
\be
[{{\tilde \sigma} \, \bar{\tilde \sigma} +\tilde{\Phi}_{00}}] = \frac{\epsilon r_H^2}{v^4}+O(v^{-5});
\ee 
where for future use we have introduced the dimensionless constant $\epsilon$ to parametrize 
the leading order term.

\subsection{Caustics in late phase}\label{tailseint}

\begin{figure}[!h]
\centering
\includegraphics[clip,width=0.48\textwidth]{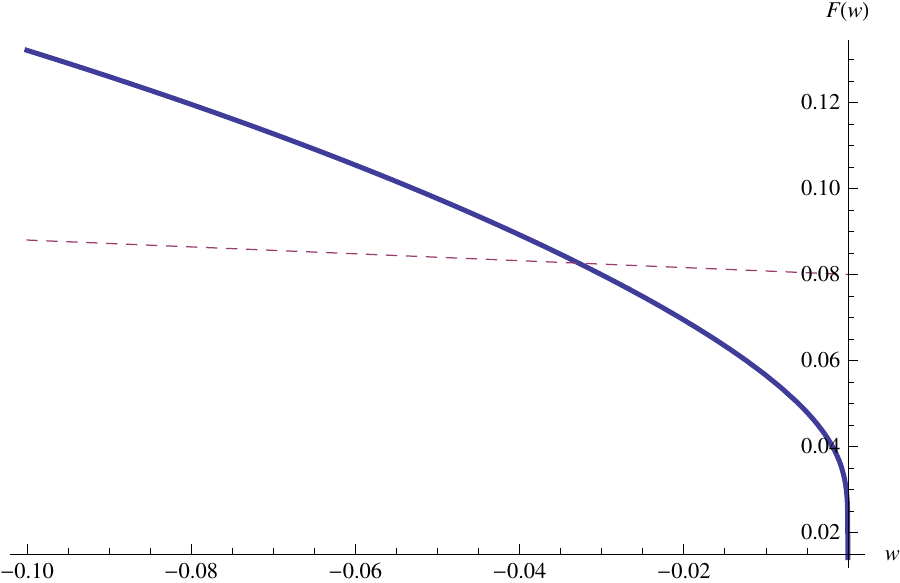}
\caption{The form of the function $F(w)$ guaranties that there exists a $w_0$ such that for $0>w>w_0$ the caustic 
free condition (\ref{cfc}) is satisfied.
The dashed line represents the function $16(1-w)/200$ which explicitly shows that there is a caustic free region 
in the case $\epsilon=200$.
All the other values of $\epsilon$ look qualitatively the same.
}
\label{f(x)}
\end{figure}

According to studies of linear perturbations of Schwarzschild geometries \cite{Gundlach94,Dafermos:2005yw}
one has that 
\be
[{{\tilde \sigma} \, \bar{\tilde \sigma} +\tilde{\Phi}_{00}}] (u\to\infty, v)= \frac{\epsilon r_H^2}{v^4} ,
\ee
where $u=t-r_*$ and $v=t+r_*$ for \be r_*=r+r_H\log\left(\frac{r-r_H}{r_H}\right)\ee
the usual tortoise coordinate, and $r_H=2M$ the radius of the horizon. From this we get that $v=u+2r_*$ hence
\be
v=u+2r+2r_H\log\left(\frac{r-r_H}{r_H}\right).
\ee
By making the same choice of region as underneath Equation \ref{prima} in the previous section,  
the caustic free condition (\ref{caustics}) becomes
\ba
&&\int\limits_{r_H (1-w)}^{\infty} \frac{\epsilon r^2_H r^2}{v^4} dr=\n \\
 && \int\limits_{r_H (1-w)}^{\infty} \frac{\epsilon r^2_H r^2}{[u+2r+2r_H\log\left(\frac{r-r_H}{r_H}\right)]^4} dr\le r_H (1-w).
\n \ea
The previous condition can be simplified by introducing the variable $x=r/r_H$, from which one gets
\be\label{cfc}
\epsilon F(w)=\epsilon \int\limits_{(1-w)}^{\infty} \frac{ x^2}{[x+\log\left(\frac{1-x}{w}\right)]^4} dx\le {16  (1-w)},
\ee
where we used that $u=-{2 r_H}\log(-w)$.  The function $F(w)$ is shown in figure \ref{f(x)}. 
It is clear from its behaviour close to $w=0$ that there is always some
$w_0$ such that there are no caustics in the region $r\in (r_H (1-w), \infty)$ for $0\le w<w_0$. 
This concludes the proof that there is a caustic free region in a neighbourhood of $i^{+}$ bounded 
by a portion of $\sI^{+}$ and the horizon $H$.

\subsection*{Acknowledgements}
We acknowledge financial support from CONICET, SeCyT-UNC, Foncyt and by 
the Agence Nationale de la Recherche; grant
ANR-06-BLAN-0050. A.P. was supported by {\em l'Institut Universitaire de France}.




\begingroup\raggedright\endgroup

\end{document}